\documentclass[twoside]{article}
\usepackage{fleqn,espcrc2}
\usepackage{epsfig}
\usepackage{fleqn}

\newcommand{\AmS}{{\protect\the\textfont2
\renewcommand{\theequation}{\thesection.\arabic{equation}}

A\kern-.1667em\lower.5ex\hbox{M}\kern-.125emS}}

\title{Non-Singular String-Cosmologies From Exact Conformal Field Theories}

\author{H.J. de Vega\address{Universit\'{e} de Paris VI et VII.
4, Place Jussieu, 75252 Paris, France.},
\underline{A.L. Larsen}\address{Department of
Physics, University of Odense,
Campusvej 55, 5230 Odense M, Denmark.}
and
N. S\'{a}nchez\address{Observatoire de Paris,
DEMIRM. 61, Avenue
de l'Observatoire, 75014 Paris, France.}}

\begin{document}

\begin{abstract}
Non-singular two and three dimensional
string cosmologies are constructed using the exact conformal
field theories corresponding to SO(2,1)/SO(1,1) and SO(2,2)/SO(2,1).
{\it All}
semi-classical curvature singularities are canceled in the exact theories for
both of these cosets, but
some new quantum curvature singularities emerge. However, considering different
patches of the global manifolds, allows the construction of non-singular
spacetimes with cosmological interpretation. In both two and three
dimensions, we construct non-singular oscillating cosmologies, non-singular
expanding and inflationary cosmologies including a de Sitter
(exponential) stage with positive scalar curvature as well as
non-singular contracting and deflationary cosmologies. Similarities between
the two and three dimensional cases suggest a general picture for higher
dimensional coset cosmologies:  Anisotropy seems to be a generic unavoidable
feature, cosmological singularities are generically avoided and it
is possible to construct non-singular cosmologies where
some spatial dimensions are experiencing
inflation while the others experience deflation.
\end{abstract}

\maketitle

\section{INTRODUCTION}

Fundamental quantum strings demand a conformally invariant background
for quantum consistency (conformal invariance is a
necessary although not sufficient condition for consistency). However, most
curved spacetimes that were
historically of physical interest in general relativity and cosmology are
not conformally invariant. For instance black hole spacetimes
and Friedman-Robertson-Walker universes,  are not of this type. In string
theory, a large amount of generalizations of these general relativity
solutions have been obtained as solutions of the renormalization group
$\beta$-function equations (low energy effective string
equations). These solutions involve, besides the metric, a number of
massless fields including the dilaton and antisymmetric tensor fields.
The $\beta$-function equations are the basis for most investigations in
string cosmology. However, they are perturbative and only known to the
lowest orders in $\alpha'$, and therefore, the corresponding solutions
are only ensured to be conformally invariant to the lowest orders in
$\alpha'$. It is not generally clear how the higher order corrections
will change these solutions, and possible non-perturbative solutions
seem to be completely missed in this framework.

A different approach to string cosmology is based on group manifolds and
coset spaces and the corresponding WZW \cite{wit1} and gauged WZW models
\cite{wit2}. These models provide new curved backgrounds that are conformally
invariant to all orders in  $1/k$ (where $k\sim1/\alpha'$ is the level of
the WZW model). In the case  of group manifolds, these models are generally too
simple to describe  physically interesting and realistic
cosmologies. This is exemplified by the well-studied $SL(2,R)$ WZW
model, which describes Anti-de Sitter space. For the coset spaces,
on the other hand, the models are generally so complicated that it is
difficult even to extract what they describe from a manifest spacetime
point of view. In fact, the exact (all orders in $1/k$) metric,
dilaton etc, have only been obtained in a limited number of
low-dimensional cases (for a rewiew of such solutions, see
Refs.\cite{ibars1,tseytlin}). And in most of these cases, the physical
spacetime properties have not really been extracted so far. Most
studies of coset space gauged WZW models still restrict themselves to lowest
order in $1/k$, and are thus essentially equivalent to the studies based on
the $\beta$-function equations.

The purpose of the present paper is to show that interesting non-trivial
and non-singular
cosmologies can be obtained from the cosets  $SO(2,1)/SO(1,1)$ and
$SO(2,2)/SO(2,1)$. Various aspects of string theory on these cosets
have been investigated in the literature, see for instance
Refs.\cite{wit1}-\cite{vafa}. Cosmological interpretations have been
considered in Refs.\cite{fradkin2,vafa,muchascosmo}, but mostly to lowest
order in $1/k$

The two dimensional coset $SO(2,1)/SO(1,1)$, to all orders in $1/k$,
has been given a cosmological interpretation in
\cite{vafa}, but only in one of the coordinate patches.

As for the three dimensional coset $SO(2,2)/SO(2,1)$, a cosmological
interpretation was attempted in \cite{fradkin2}, but only in certain
coordinate patches and only to first order in $1/k$. The cosmologies
obtained in \cite{fradkin2}, however, are completely changed in the exact
theory.

Interestingly enough, quantum corrections not only cancel the
semiclassical singularities but also change the sign of the scalar
curvature for oscillatory cosmologies. In non-oscillating
cosmologies, the scalar curvature evolves from positive for large and
negatives times $ t $, to negative near $t = 0$.

\section{TWO-DIMENSIONAL COSET}

The $SO(2,1)/SO(1,1)$ metric and dilaton, to all orders in $1/k$, are given by
\cite{verlinde}:
\begin{equation}
ds^2=2(k-2)\left
[ \frac{db^2}{4(b^2-1)}-\beta(b)\; \frac{b-1}{b+1}\;dx^2\right]
\end{equation}
\begin{equation}
\Phi(b)=\ln\frac{b+1}{\sqrt{\beta(b)}}+ \mbox{const.}
\end{equation}
where:
\begin{equation}
\beta(b)\equiv\left( 1-\frac{2}{k}\;\frac{b-1}{b+1}\right)^{-1}
\end{equation}
In these $(b,x)$-coordinates, the global manifold is
\begin{eqnarray}
-\infty<b<\infty,\;\;\;\;-\infty<x<\infty\nonumber\; .
\end{eqnarray}
The scalar curvature is given by:
\begin{equation}
R(b)=\frac{4}{k-2}\;\frac{k(k-4)+k(k-2)b}{[k+2+(k-2)b]^2}
\end{equation}
We see here the presence of new quantum curvature singularities at
\begin{equation}
b=-\frac{k+2}{k-2}
\end{equation}
plus coordinate singularities (horizons) at $b=\pm 1$.

In the semi-classical
limit $(k\rightarrow \infty)$, the scalar curvature reduces to
\begin{equation}
R(b) \buildrel{k \to \infty}\over= \frac{4}{k}\; \frac{1}{1+b}
\end{equation}
Comparison of (4) and (6) shows that in the semi-classical
limit the quantum curvature singularities coalesce with the coordinate
singularities (horizons) at $b=\pm 1$.

The fact that the semi-classical curvature singularities become reduced to
coordinate singularities, and that new curvature singularities appear in the
exact (all orders in $1/k$) theory, will turn out to hold also for the
3-dimensional cosets, to be discussed in the next section.

For conformal invariance we demand that
\begin{equation}
C\left(\frac{SO(2,1)}{SO(1,1)}\right)=26
\end{equation}
leading to $k=9/4$. Thus the curvature singularity (5)
is located at $b=-17$, that
is to say, in the coordinate patch to the left of the left horizon in the
$(b-x)$-coordinate system. Clearly, it is then possible to construct
non-singular
spacetimes by considering the other coordinate patches. In the present case, we
are interested in constructing simple cosmological spacetimes, by which we
mean,
spacetimes with signature $(-+)$ and line-element of the form
\begin{equation}
ds^2=-dt^2+A^2(t)\; dx^2
\end{equation}
where $t$ plays the role of cosmic time, $x$ is the spatial coordinate and
$A(t)$ is the scale factor.

\subsection{Oscillating Cosmology}

Considering the coordinate patch between the two horizons, $|b|\leq 1$, and
using the parametrization $ b=\cos 2 t $, we get the cosmology \cite{vafa}
\begin{equation}
ds^2=\frac{1}{2}\left( -dt^2+\frac{\tan^2 t }{1+\frac{8}{9}\tan^2 t }\;dx^2
\right)
\end{equation}
\begin{equation}
\Phi(t)=\ln\left( \cos^2 t\; \sqrt{1+\frac{8}{9}\tan^2 t
}\;\right)+\mbox{const.}
\end{equation}
The scalar curvature
\begin{equation}
R(t)=-72\; \frac{4-\cos^2 t }{(8+\cos^2 t )^2}
\end{equation}
is oscillating, finite and always negative
\begin{eqnarray}
&R(t)& \hspace{-6.5mm} \in[-\frac{9}{2},\;-\frac{8}{3}] \nonumber \\
&<R(t)>&=-\frac{5}{\sqrt{2}}
\approx -3.53..
\end{eqnarray}
It is an interesting observation that the quantum corrections not only
cancel the semi-classical curvature singularities, but also change the
sign of the scalar curvature, since semi-classically ($k\rightarrow\infty$)
\begin{equation}
R(t)\rightarrow\frac{4}{k}\;\frac{1}{1+\cos 2t }
\end{equation}
which is always positive (and sometimes positive infinity).
We shall see in the next section that this feature also holds in the
3-dimensional case.

Finally, we make
a few comments about the dilaton (10). In the present notation and
conventions, the string-coupling $g_s$ is given by
\begin{equation}
g_s=e^{-\Phi/2}=\left(\cos^2 t \; \sqrt{1+\frac{8}{9}\tan^2 t }\;\right)^{-1/2}
\end{equation}
up to an arbitrary positive multiplicative constant. It follows that we should
strictly speaking only trust the cosmological solution in the regions near
$t=0$, $t=\pm\pi$, $t=\pm 2\pi$ etc. That is to say, in the regions where
the scale factor is small. In the other
regions where the scale factor is large, we
should expect that string-loop corrections will be important and possibly
change the solution dramatically.  Thus, concentrating on (say) the region
around $t=0$, the line-element (9) describes a universe experiencing
deflationary contraction (negative $t$) followed by
deflationary expansion (positive $t$). And always with negative scalar
curvature.

\subsection{Deflationary Cosmology}

The oscillating cosmology (9) was obtained for the value $k=9/4$,
corresponding
to conformal invariance. However, using other values of $k$, it is possible
to construct other types of non-singular cosmologies. In that case,
conformal invariance must be ensured by adding other conformal field
theories. Moreover, problems with unitarity will possibly appear in such cases.
In this subsection, we shall not deal with these problems, but instead
concentrate on the possible    cosmologies that can be obtained by keeping $k$
arbitrary.

Taking $k=-|k|<0$, we observe that the curvature singularity (5)
is located between the two horizons $b=\pm 1$. It is then possible to construct
non-singular spacetimes by specializing to the two patches outside the
horizons.

Consider first the patch $b\geq 1$ using the parametrization
$b=\cosh 2t $. Then Eqs.(1)-(3) become
\begin{equation}
ds^2=2(2+|k|)\left( -dt^2+\frac{\tanh^2 t }{1+\frac{2}{|k|}\tanh^2 t }
\; dx^2 \right)
\end{equation}
\begin{equation}
\Phi(t)=\ln\left(\cosh^2 t  \;\sqrt{1+\frac{2}{|k|}\tanh^2 t }\;\right)+
\mbox{const.}
\end{equation}
Notice that it is crucial to include the overall prefactor $2(k-2)$ in
the line-element to get the desired signature.

In this case the scalar curvature is given by
\begin{equation}
R(t)=-\frac{2|k|}{2+|k|}\; \frac{(2+|k|)\cosh^2 t +1}{[(2+|k|)\cosh^2 t
-2]^2}
\end{equation}
while the Hubble function and its derivative become
\begin{equation}
H=\frac{\dot{A} (t) }{A(t)}=
\frac{\cosh t }{(\cosh^2 t +\frac{2}{|k|}\sinh^2 t ) \sinh t }
\end{equation}
\begin{equation}
\dot{H}=-|k|\; \frac{(2+|k|)(1+2\sinh^2 t)\cosh^2 t -2}
{[(2+|k|)\cosh^2 t -2]^2\sinh^2 t }
\end{equation}
It follows that the universe has negative scalar curvature and it is always
deflationary ($\dot{H}<0$). More precisely, starting from flat Minkowski
space $(t=-\infty)$, the universe experiences deflationary contraction until
$t=0$. Still deflationary, it then expands and approaches flat Minkowski
space again for $t\rightarrow\infty$; see Figure 1.

\begin{figure}
\centerline{\epsfig{figure=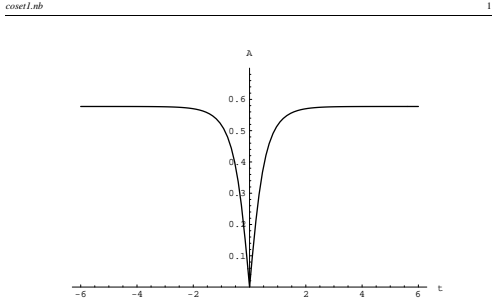, height=4cm,angle=0}}
\caption {The scale factor (15) as a function of time. Here shown for $k=-1$.
This cosmology is always deflationary.}
\end{figure}

The string-coupling in this case is given by
\begin{equation}
g_s=e^{-\Phi/2}=\left( \cosh^2 t  \;\sqrt{1+\frac{2}{|k|}\tanh^2 t }\;
\right)^{-1/2}
\end{equation}
which is always finite, and in fact goes to zero for $t\rightarrow\pm\infty$;
see Figure 2.
Thus, this non-singular cosmology should be trusted everywhere.

\begin{figure}
\centerline{\epsfig{figure=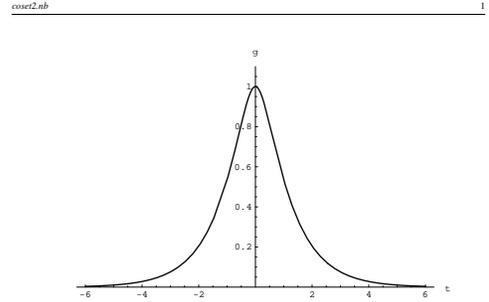, height=4cm,angle=0}}
\caption {The string-coupling (20) corresponding to Figure 1.
 The string-coupling is finite for all $t$.}
\end{figure}

\subsection{Inflationary Cosmology}

A more interesting cosmology can be constructed by considering
the patch $b\leq -1$ (and still $k<0$) instead of $b\geq 1$.
Using the parametrization
$b=-\cosh 2t $, Eqs.(1)-(3) become
\begin{equation}
ds^2=2(2+|k|)\left( -dt^2+\frac{\coth^2 t }
{1+\frac{2}{|k|}\coth^2 t } \;dx^2\right)
\end{equation}
\begin{equation}
\Phi(t)=\ln\left(\sinh^2 t \; \sqrt{1+\frac{2}{|k|}\coth^2 t }\;\right)+
\mbox{const.}
\end{equation}
The scalar curvature is now given by
\begin{equation}
R(t)=\frac{2|k|}{2+|k|}\; \frac{(2+|k|)\sinh^2 t
-1}{[(2+|k|)\sinh^2 t +2]^2}
\end{equation}
while the Hubble function and its derivative become
\begin{equation}
H=-\frac{\sinh t }{(\sinh^2 t +\frac{2}{|k|}\cosh^2 t )\cosh t }
\end{equation}
\begin{equation}
\dot{H}=|k|\;\frac{(2+|k|)(1+2\sinh^2 t )\sinh^2 t -2}{[(2+|k|)
\sinh^2 t +2]^2\cosh^2 t }
\end{equation}
Now the situation is a little more complicated than in the $b\geq 1$ patch.
However, notice that
\begin{equation}
R(t)>0\;\;\;\;\Leftrightarrow\;\;\;\;\cosh^2 t  > \frac{3+|k|}{2+|k|}
\end{equation}
as well as
\begin{equation}
\dot{H}>0 \Leftrightarrow \cosh^2 t  >
\frac{1}{4}\left( 3+\sqrt{1+\frac{16}{2+|k|}}\;\right)
\end{equation}
Thus the universe starts out as flat Minkowski space at $t=-\infty$. It then
experiences inflationary $(\dot{H}>0)$
expansion with positive scalar curvature. Just before
the scale factor reaches its maximal value at $t=0$, the scalar curvature
becomes negative and the
expansion becomes deflationary. For $t>0$, the evolution is simply
time-reversed; See Figure 3.

\begin{figure}
\centerline{\epsfig{figure=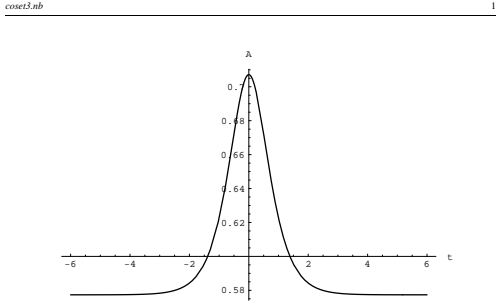, height=4cm,angle=0}}
\caption {The scale factor (21) as a function of time. Here shown for $k=-1$.
This cosmology is expanding and inflationary for large negative $t$.}
\end{figure}

It is particularly interesting that the universe is inflationary ($\dot{H}>0$)
with positive scalar curvature for (large) negative $t$. This suggests that
the universe goes through some kind of de Sitter phase. This is confirmed by
expanding the scale factor for large negative $t$ ($t<<0$)
\begin{eqnarray}
A(t)&=&\sqrt\frac{\coth^2 t }{1+\frac{2}{|k|}\coth^2 t }\\
&\approx &
\sqrt{\frac{|k|}{2+|k|}}\left(1+\frac{2|k|}{2+|k|}\;e^{2t}\right)
\end{eqnarray}
thus there is in fact an element of exponential expansion.

Finally, let us consider the string-coupling also in this case. It is given by
\begin{equation}
g_s =e^{-\Phi/2}=\left(\sinh^2 t  \;\sqrt{1+\frac{2}{|k|}\coth^2 t
}\;\right)^{-1/2}
\end{equation}
which is finite everywhere except near $t=0$; see Figure 4. In particular, the
string-coupling is finite during the phase of inflationary expansion.

\begin{figure}
\centerline{\epsfig{figure=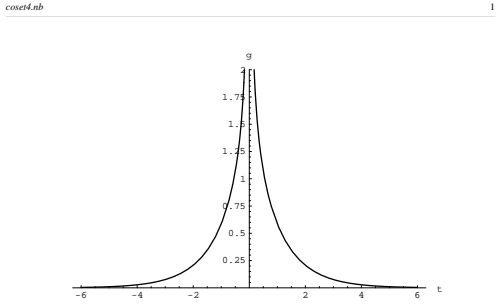, height=4cm,angle=0}}
\caption {The string-coupling (30) corresponding to Figure 3.
The string-coupling is finite everywhere except near $t=0$.}
\end{figure}

\section{THREE-DIMENSIONAL COSET}

The computations are now going to be somewhat more complicated. Here we
shall only
present the results; the detailed computations can be found in \cite{all}.

The $SO(2,2)/SO(2,1)$ metric and dilaton, to all orders in
$1/k$, are given by
\cite{bars2}
\begin{eqnarray}
ds^2=2(k-2) [
\frac{db^2}{4(b^2-1)}&+&g_{uu} \; du^2+g_{vv} \; dv^2 \nonumber \\
 &+&2 \;g_{uv} \; du \; dv ]
\end{eqnarray}
\begin{equation}
\Phi(b,u,v)=\ln\frac{(b^2-1)(v-u-2)}{\sqrt{\beta(b,u,v)}}+\mbox{const.}
\end{equation}
where the explicit expressions for $g_{uu}$, $g_{vv}$, $g_{uv}$ and $\beta$
can be
found in \cite{bars2}.

In these $(b,u,v)$-coordinates, the global manifold for $SO(2,2)/SO(2,1)$ is
\cite{bars1,bars2}
\begin{eqnarray}
&b^2>1&\;\;\;\;\mbox{and}\;\;\;\;uv>0,\nonumber\\
&b^2<1&\;\;\;\;\mbox{and}\;\;\;\;uv<0,
\nonumber
\end{eqnarray}
excluding $0<v<u+2<2$ in the second case. However, other regions in the
$(b,u,v)$-space can be reached by going for instance to
the $SO(3,1)/SO(2,1)$ coset, for which the metric and dilaton are
formally the same as above. For more details, see Refs.\cite{bars1,bars2}.

The general expression for the scalar curvature, corresponding to the metric
(31), is quite complicated. It is most conveniently written as a fourth
order
polynomium in $b$ divided by the square of a second order polynomium in $b$:
\begin{equation}
R=\frac{4}{k-2}\; \frac{c_0+c_1b+c_2b^2+c_3b^3+c_4b^4}
{(d_0+d_1b+d_2b^2)^2}
\end{equation}
where the coefficients $c_{i}$ and $d_{j}$ are functions of
the remaining coordinates $(u,v)$ and the parameter $k$
\begin{eqnarray}
c_{i}&=&c_{i}(u,v;k),\;\;\;\;i=0,1,2,3,4\nonumber\\
d_{j}&=&d_{j}(u,v;k),\;\;\;\;j=0,1,2\nonumber
\end{eqnarray}
Their explicit expressions are given in \cite{all}.

Comparison of the semi-classical expression ($k\rightarrow \infty$)
and the exact  expression (33) for the
scalar curvature shows that {\it all} semi-classical curvature singularities
($b=1$, $b=-1$ and $v=u+2$) have disappeared, but that new quantum curvature
singularities have appeared. The situation is thus completely analogous to the
2-dimensional case: The semi-classical curvature singularities become
coordinate singularities (horizons), but new quantum curvature singularities
have appeared elsewhere. The curvature singularities in the exact geometry
correspond to the zeroes of the denominator in (33).
They precisely
correspond to the singularities of the function $\beta(b,u,v)$.

By considering different values of $k$, one can move the singularity
surface around in the global manifold. It is then possible,
as in the 2-dimensional
case, to construct non-singular spacetimes by considering a single
coordinate patch. Notice that for conformal invariance, we should demand
that
\begin{equation}
C\left(\frac{SO(2,2)}{SO(2,1)}\right) = 26
\end{equation}
leading to $k=(39\pm 5\sqrt{13})/23$.

\subsection{Oscillating Cosmology}

Consider first the case where $k=(39+5\sqrt{13})/23\approx 2.48..\;$. It is
then
possible to construct a cosmology in the coordinate patch
\begin{equation}
|b|\leq 1,\;\;\;\;u\geq 0,\;\;\;\;v\leq 0\nonumber
\end{equation}
using the parametrization
\begin{equation}
b=\cos 2t\; ,\quad u=2x^2\; , \quad v=-2y^2 \nonumber
\end{equation}

It is easy to see that this is a non-singular cosmology with $t$
playing the role of cosmic time, $(x,y)$ are the spatial coordinates and
the signature is $(-++)$, as it should be. The cosmology is however somewhat
complicated: It is periodic in time, but non-homogeneous and highly
non-isotropic.
In fact, the `scale factors' for the two spatial directions are
oscillating with a phase difference of $\pi/2$.

To gain a little more insight into this cosmology, we consider the
region near the spatial `origin' $(x,y)=(0,0)$. The scalar curvature reduces
to
\begin{equation}
R(t,0,0)=\frac{1}{k-2}\; \frac{\tilde{c}_0+\tilde{c}_2\cos^2 2t +
\tilde{c}_4\cos^4 2t }{[k^2-(k-2)^2\cos^2 2t ]^2}
\end{equation}
where the coefficients $(\tilde{c}_0,\tilde{c}_2,\tilde{c}_4)$ are given
in \cite{all}. It follows that
\begin{equation}
R(t,0,0)\in[-3.1435..,\;-2.2988..]
\end{equation}
as well as
\begin{equation}
<R(t,0,0)>\approx -2.7126..
\end{equation}
where the numerical values were obtained for $k=(39+5\sqrt{13})/23$.
As in the 2-dimensional case, we notice therefore
that the quantum corrections changed the
sign of the curvature, since semi-classically $(k\rightarrow\infty)$
\begin{equation}
R(t,0,0)\buildrel{k \to \infty}\over=
\frac{4}{k} \; \frac{3+\cos^2 t }{2\sin^2 t }
\end{equation}
which is always positive (and sometimes positive infinity).

In the region near $(x,y)=(0,0)$, one can further define scale factors
\begin{eqnarray}
A(t)&=&\sqrt{g_{xx}(t,0,0)}\\
B(t)&=&\sqrt{g_{yy}(t,0,0)}
\end{eqnarray}
both of which oscillate between $0$ and $\sqrt{k-1}\approx 1.216..$
with a phase difference of $\pi/2$.

Finally, let us return to the dilaton (for generic $(t,x,y)$). The
string-coupling is given by
\begin{equation}
g_s =\left[\frac{(x^2+y^2+1)\cos^2 t \; \sin^2 t }
{\sqrt{\beta(t,x,y)}}\right]^{-1/2}
\end{equation}
Thus the string-coupling blows up at $t=0$, $t=\pm\pi/2$, $t=\pm \pi$ etc. It
means that we should only trust the solution in the intermediate regions
where $ g_s $ is not large.

\subsection{Non-Oscillating Cosmologies}

In the previous subsection we obtained an oscillating cosmology using the
value $k=(39+5\sqrt{13})/23\approx 2.48..$, corresponding to conformal
invariance. Most formulas were however presented keeping $k$ arbitrary, and
the oscillating cosmologies in fact exist for arbitrary $k>2$. In this
subsection we shall show that it is
possible to construct non-singular 3-dimensional
non-oscillating
cosmologies when $k<1$. Interestingly enough, the condition (34) of conformal
invariance gives rise to the possibility $k=(39-5\sqrt{13})/23\approx 0.912..$,
but in the following we just keep $k$ arbitrary but less than $1$. Possible
problems with unitarity will not be dealt with here.

Thus we take $k<1$ and consider first the patch
\begin{equation}
b\geq 1,\;\;\;\;u\geq 0,\;\;\;\;v\leq 0
\end{equation}
and use the parametrization
\begin{equation}
b=\cosh 2t\; ,\quad u=2x^2\; ,\quad v=-2y^2
\end{equation}
Actually the patch (44) is not part of the global manifold for $SO(2,2)/
SO(2,1)$, so one has to go to the de Sitter coset $SO(3,1)/SO(2,1)$
instead
\cite{bars1,bars2}.
The general expressions for the metric and dilaton (31)-(32)
are however unchanged.

Using the parametrization (45), the new metric and dilaton are similar to the
ones of Subsection 3.1. except that trigonometric functions
become hyperbolic functions
and $k-1$ becomes $1-k$. Since we now consider the case
where $k<1$, it is then clear that it describes a cosmology with
the
correct signature $(-++)$. Furthermore, it is easily seen that the
cosmology is non-singular. The change from trigonometric functions to
hyperbolic functions obviously has dramatic consequences. The cosmology
is however
still non-isotropic and non-homogeneous, but it is no longer oscillating.
In fact, the time-dependence of the metric actually disappears for
$t\rightarrow\pm\infty$, i.e., the universe becomes
static in these limits (however,
the dilaton stays time-dependent). More precisely, both `scale factors'
start out with constant values at $t=-\infty$. Then one of them increases
monotonically towards a maximal value, while the other one decreases
monotonically to zero for $t\rightarrow 0_-$. For $t>0$, their behaviour is
simply time-reversed.

As for the oscillating cosmology in the previous subsection, let us define
scale factors analogous to (41)-(42). In the present case the results are
shown in
Figure 5. A careful analysis of the corresponding Hubble functions and their
derivatives shows that for $t<<0$, one scale factor is contracting and
deflationary
while the other scale factor is expanding and inflationary. This is most easily
seen from the expansions
\begin{equation}
A(t)\approx\sqrt{\frac{1-k}{2-k}}\left(1-2 \;\frac{1-k}{2-k}\;e^{2t}\right),
\;\;t<<0
\end{equation}
\begin{equation}
B(t)\approx\sqrt{\frac{1-k}{2-k}}\left(1+2 \;\frac{1-k}{2-k}\;e^{2t}\right),
\;\;t<<0
\end{equation}
and still taking into account that $k<1$.

\begin{figure}
\centerline{\epsfig{figure=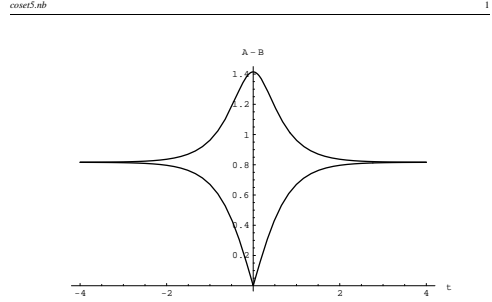, height=4cm,angle=0}}
\caption {The two scale factors of Subsection 3.2. as a function of time. Here
shown for
$k=-1$. For large negative $t$, one scale factor is expanding and inflationary
while the other scale factor is contracting and deflationary.}
\end{figure}

It is interesting that the scale
factor $A(t)$ is very similar to the scale factor
of the deflationary 2-dimensional cosmology of Subsection 2.2, while the
scale factor $B(t)$ is very similar to the scale factor of the inflationary
2-dimensional cosmology of Subsection 2.3. In particular, the scale factor
$B(t)$ has en element of exponential expansion as seen from Eq.(47).
In that sense, the 2-dimensional
coset comprises all the main features of the 3-dimensional coset.

The string-coupling in this case is shown in Figure 6.
It is finite everywhere except near $t=0$. Asymptotically $(t\rightarrow
\pm\infty)$ it goes to zero.
Thus the solution should be trusted everywhere
except near $t=0$.

It is also possible to construct a non-singular non-oscillating
cosmology in the patch
\begin{equation}
b\leq -1,\;\;\;\;u\geq 0,\;\;\;\;v\leq 0
\end{equation}
using the parametrization
\begin{equation}
b=-\cosh 2t\; ,\quad u=2x^2\; ,\quad v=-2y^2
\end{equation}
and still taking $k<1$ (also in this case one has to go to the de Sitter
coset $SO(3,1)/SO(2,2)$, see Ref.\cite{bars1,bars2}). However, the resulting
cosmology is identical with the previous one, up to an interchange of $x$
and $y$.

\begin{figure}
\centerline{\epsfig{figure=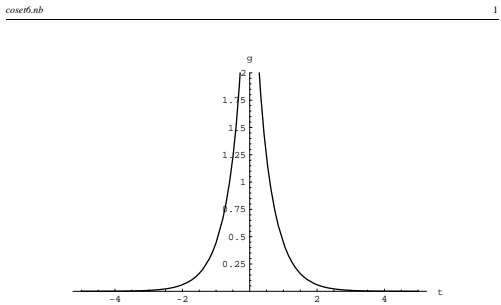, height=4cm,angle=0}}
\caption {The string-coupling corresponding to Figure 5.
The string-coupling is finite everywhere except near  $t=0$.}
\end{figure}

\section{CONCLUSIONS}

The cosmological geometries found in this paper are exact conformal
field theories. That is, they describe string vacua. Physically, these
are spacetimes where the (massless)
dilaton and graviton fields are present. We find that
 such manifolds provide {\bf non-singular} spacetimes with
cosmological interpretation. These are of two types: oscillating and
non-oscillating cosmologies. The non-oscillating metrics start as
Minkowski spacetimes for early times, evolve through inflationary
expansion (with {\it positive scalar curvature}), then  pass through a
deflationary contraction ending as a Minkowski spacetime for late
times.

The coset string cosmologies studied in this paper correspond to
analogies  of de Sitter spacetime in string vacua. In General
Relativity, the global de Sitter manifold describes a contracting
phase (for $ t \leq 0 $) and then an expanding universe for $ t \geq 0 $. Only
the expanding patch describes the physical space. Similarly, in string
cosmologies one should not consider the physical space as the whole
global manifold but only a part of it.

\newpage

\end{document}